# Hydrogen sensing by sol-gel grown NiO and NiO:Li thin films

I. Sta[*,1], M. Jlassi[1], M. Kandyla[3], M. Hajji[1,2], P. Koralli[3], R. Allagui[1], M. Kompitsas[3], and H. Ezzaouia[1]

[1] *Photovoltaic Laboratory, Research and Technology Centre of Energy, Borj-Cedria Science and Technology Park, BP 95, 2050 Hammam-Lif, Tunisia*

[2] *Ecole Nationale d'Electronique et des Communications de Sfax, Technopôle de Sfax, Route de Tunis Km 10, Cité El Ons, B.P. 1163, 3021 Sfax, Tunisia*

[3] *National Hellenic Research Foundation, Theoretical and Physical Chemistry Institute, 48 Vasileos Konstantinou Ave., 11635 Athens, Greece*

[*]Email of corresponding author: imenstalpv@yahoo.fr

## Abstract

Hydrogen sensors have been prepared using nickel oxide (NiO) and lithium-doped nickel oxide (NiO:Li) thin films, deposited on glass substrates by the sol-gel spin coating technique. The surface morphology, structure, optical and electrical properties of the obtained films were studied. Hydrogen sensing results are presented for three operating temperatures (140, 160, and 180 °C) and for hydrogen concentrations ranging from 1000 to 15000 ppm in synthetic air. The NiO and NiO:Li (2% and 8% doping concentrations) sensors show maximum responses for the operating temperature of 180 °C. When tested at different hydrogen concentrations in air, the lithium-doped NiO sensors showed a higher response than the undoped NiO films.

# 1. Introduction

Hydrogen is an abundant, efficient, renewable, clean energy source. It is currently utilized in transportation, industrial, and residential applications, under different temperature and pressure conditions [1, 2]. For the safe production and distribution of hydrogen on a large scale, sensitive sensing devices with reliable operation are essential. In order to prevent explosions, it is important to develop devices that detect the presence of hydrogen much below the lowest explosion limit (LEL) of 40,000 ppm in air [3].

Over the past decades, thin-film gas sensors have been developed based on a variety of materials, including metal oxide semiconductors [4], conducting polymers [5, 6], organic-inorganic hybrids [7], and graphene [8], among others. Gas sensors based on metal-oxide semiconducting films are widely studied for the detection of toxic and explosive gases. Metal oxides with n-type conductivity have been widely studied as gas sensors [9], in contrast with p-type metal oxides, which are less common [10]. Nickel oxide (NiO) is one of the few p-type metal-oxide semiconductors [11]. It is an antiferromagnetic, electrochromic [12, 13], and catalytic [14, 15] material with excellent chemical stability [16], high transparency and electrical conductivity [17], showing promising gas sensing properties [18 - 21]. Thin-film sensors, based on p-type NiO, have been developed by different methods, such as pulsed laser deposition [22, 23], dip coating [24], chemical spray pyrolysis [25], and sputtering [18, 26], for a variety of analyte gases.

Thin NiO films, deposited by magnetron sputtering, were able to detect 500 ppm of hydrogen at an operating temperature of 200 - 500 °C [26]. Fasaki *et al.* reported hydrogen detection for concentrations ranging from 10 000 to 500 ppm at 185 - 400 °C [22] and Stamataki *et al.* reported the detection of 30 000 ppm of hydrogen at 80 - 120 °C [23], employing pulsed laser-deposited NiO films. Wilches *et al.* employed NiO films, developed by chemical spray deposition, for the detection of hydrogen in a high concentration range from 3000 to 30 000 ppm at 300 °C [25]. Steinebach *et al.* reported that a sputter-deposited NiO sensor showed a response up to 50%, for hydrogen concentration levels of 500 - 10,000 ppm at an elevated operating temperature of 600 °C [27]. Du *et al.* investigated the effect of NiO particle size, synthesized by chemical methods, on the response to 1000 ppm of hydrogen at 150 °C [28]. Soleimanpour *et al.* detected 3000 ppm of hydrogen with high sensitivity at 175 °C, for NiO thin films produced by a sol–gel method [29]. Recently, lithium has been used as a dopant in order to improve certain properties of NiO, such as its transparency and electrical conductivity [30-32], and its behavior as a thermoelectric [33] and electrochromic [34] material. As a hydrogen sensor, lithium-doped NiO (NiO:Li) was studied by Shin *et al.* [35], Gardun͂o-Wilches *et al.* [25], and Matsumiya [36]. NiO:Li films, prepared by the sol gel method, have not been investigated as resistive hydrogen sensors yet.

In this work, we prepared undoped NiO and NiO:Li doped thin films by the sol-gel spin coating technique, to be studied as resistive hydrogen sensors with low operating temperature and detection limit. The sol–gel method is able to produce large-area, homogeneous thin films with a controlled composition. Furthermore, it does not require the use of vacuum, as sputtering and pulsed laser deposition. We find that the maximum response to hydrogen was obtained for NiO and NiO:Li films at an operating temperature of 180 °C. Additionally, for all hydrogen concentrations and operating temperatures employed, the response of the NiO:Li films is better than the response of NiO.

## 2. Experimental

*2.1. Film preparation and processing*

The precursor solutions for both undoped and Li-doped NiO were prepared by the sol–gel method, using nickel acetate tetrahydrate powder ($C_4H_6NiO_4·4H_2O$) (98%, Sigma Aldrich) and lithium chloride (LiCl) (99.0%, Sigma Aldrich) as the starting materials. Nickel acetate was dissolved in 2-methoxyethanol ($C_3H_8O_2$) (99.8%, Sigma Aldrich) and diethanolamine

(MEA) (99%, Merck) was added slowly in the solution under magnetic stirring. The molar ratio of diethanolamine to nickel acetate was maintained to 1.0 with the concentration of nickel acetate in the precursor solution being 0.5 M. The resultant solution was stirred at 80 °C for 1 h to yield a clear and homogeneous solution. Lithium chloride (as a dopant precursor) was also dissolved in 2-methoxyethanol under air atmosphere. Finally, both precursor solutions were mixed, with different volume fractions, to prepare stoichiometric, transparent, and stable nickel acetate solutions with lithium contents of 0, 2%, 4%, 6%, 8%, and 10%. The percentage of lithium in the solution was defined as the ratio between the concentration of lithium ions ($Li^+$) and the total concentration of cations ($Li^+$ and $Ni^{2+}$).

Undoped and Li-doped NiO thin films were deposited on glass substrates by spin coating, with a rate of 3000 rpm for 30 s at room temperature. Before deposition, the glass substrates were cleaned ultrasonically in acetone, ethanol, and deionized water. After each coating, the films were heated at 300 °C for 5 min in air for the solvent and organic residues to evaporate. The procedure from coating to drying was repeated four times in order to obtain films with optimized thickness with respect to their optical, morphological, and electrical properties [31]. The undoped NiO and NiO:Li thin films were deposited using the same optimized experimental conditions. Finally, all the films were annealed at 600 °C for 1 h in air, in order to enhance their electrical conductivity. Annealing at higher temperatures was avoided, in order to prevent the damaging of the glass substrates. Among the deposited NiO:Li thin films, the ones with 2% and 8% lithium concentration were chosen to be tested as hydrogen sensors, because they combine large surface area with high electric conductivity.

*2.2. Film characterization*

The film structure was determined by X-ray diffraction (XRD, Bruker D8 advance) with Cu $K_α$ radiation (λ = 1.5405 Å) in a (2θ) scanning range between 20° and 90°. An Atomic Force Microscope (AFM) (Nanoscope III) in tapping configuration was used to scan an area of 5 μm × 5 μm, in order to investigate the topography and the surface morphology of the films. The obtained results were used to estimate the surface roughness and the grain distribution of the films. Optical transmittance spectra were recorded with an UV–Vis–NIR spectrophotometer (Lambda 950), equipped with an integrating sphere, in the wavelength range from 250 to 2500 nm. The electrical resistivity was measured at room temperature by using a four-point probe method (S 302 - X). All measurements were carried out at room temperature.

*2.3. Gas sensing*

Hydrogen sensing tests were performed in a home-built sensing setup [36]. An aluminum vacuum chamber was initially evacuated to 1 Pa, and then filled with dry air/hydrogen mixtures at atmospheric pressure. The hydrogen concentration was calculated based on the partial pressures of hydrogen and air inside the chamber, as measured by a Baratron gauge [21]. The undoped and Li-doped NiO thin films were tested for hydrogen concentrations of 15 000, 10 000, 5000, 3000, 2000, and 1000 ppm in air.

The films were placed on a resistively heated stainless steel plate and the operating temperature in the 140 – 180 °C range was measured by a Ni-Cr thermocouple, mounted on the heated plate. The electric current through the films was measured with a model 485 Keithley picoamperemeter, under a constant bias voltage of 1 V, employing two copper electrodes in direct contact with the films. Electric current changes due to hydrogen sensing were digitally recorded and displayed in real time, for various operating temperatures and hydrogen concentrations. The sensor response, *S*, was calculated as:

$$S = (R_g - R_o)/R_o \qquad (1),$$

where $R_g$ and $R_o$ is the electric resistance of the sensors in the presence and in the absence of hydrogen, respectively.

# 3. Results and discussion

*3.1. Structural properties*

The X-ray diffraction patterns of pure NiO and NiO:Li films with lithium concentrations of 2% and 8% are shown in Fig. 1. For all samples, the XRD patterns show three peaks at 37.31°, 43.31°, and 62.81°, corresponding to the (111), (200), and (220) planes of the cubic structure of NiO, respectively, according to the JCPDS file no. 4-0835. The (200) peak has the highest intensity, indicating that (200) is the preferred orientation. The X-ray diffraction patterns of the films in the 2θ range of 35° to 45° are also shown in Fig.1. Doping with 8% lithium results in an increase of the intensity of the XRD peaks, indicating the crystallinity of the films improves. Additionally, the (200) peak becomes wider (full width at half maximum), suggesting that the doping process leads to a decrease in the grain size. No peaks other than the NiO peaks are observed, which implies that lithium atoms may be incorporated in the NiO lattice. The cubic phase of the NiO matrix remains unaffected by the incorporation of lithium.

The average crystallite size, *D*, was estimated by the Scherrer equation [38]:

$$D = \frac{K\lambda}{\beta \cos\theta} \qquad (2),$$

applied to the (200) peak in the XRD data, where $\lambda$ is the X-ray wavelength of 1.54 Å, $\theta$ is the Bragg angle of the XRD peak, *K* is a shape factor with a typical value of 0.9, and $\beta$ is the full width at half maximum of the peak. The values of the average crystallite sizes were found to be 21 nm, 19 nm, and 18 nm for the undoped NiO film, the NiO:Li 2% film, and the NiO:Li 8% film, respectively (Table1). We note that the crystallite size decreases with increasing dopant concentration, which implies that the smaller radius of the embedded $Li^+$ ions results in a decrease of the lattice constant [39].

*3.2. Morphological properties*

The surface morphology of undoped NiO and NiO:Li thin films was characterized by Atomic Force Microscopy. The obtained three-dimensional (3D) AFM images are shown in Fig. 2. When the lithium content increases from zero to 2%, the grains of the sample become sharper. The grain size becomes smaller, which can be related to the decrease of the crystallite size. As the lithium content increases to 8%, more of the bigger pyramids form, which increases the active surface of the film. The AFM 3D images show that lithium doping affects the density and the size of the NiO grains significantly. To determine the surface density of the samples and their grain distribution, the grains have been sorted according to their size. Figure 3 shows the obtained histogram of the grain size distribution as a function of the lithium doping of the samples. For a scanning area of 5 μm × 5 μm, the root-mean-square (RMS) surface roughness is determined as 3.03, 4.96, and 13.5 nm for the NiO and NiO:Li films with doping concentrations of 2% and 8%, respectively. The surface roughness increases with the lithium concentration.

*3.3. Optical properties*

We studied the effect of the dopant concentration on the optical transmittance and the band gap, $E_g$, of the NiO films. From Fig. 4, we observe that all films are highly transparent in the visible and near infrared parts of the electromagnetic spectrum. The average transmission value of the 2% NiO:Li film, which is the most transparent, is 83.5% for the visible wavelength range. The optical band gap of NiO and NiO:Li was estimated by employing the Tauc model [40]:

$$\alpha h\nu = A\,(h\nu - E_g)^{1/2} \qquad (3),$$

where *α* is the absorption coefficient of the film, *hv* is the photon energy, *A* is a constant, and $E_g$ is the optical band gap.

The optical band gap of the thin films was determined by extrapolating the linear section of $(\alpha h\nu)^2$ to the energy axis, as shown at the inset of Fig. 4. Table 1 lists the extrapolated band gap values for all samples, which show an increase of the band gap from 3.88 to 3.97 eV with the concentration of lithium in the NiO:Li films, which is related to the grain size increase [41]. We note that the optical band gap of NiO:Li films increases compared to that of undoped NiO samples and then it varies slightly with the increase of the lithium concentration. The obtained values of the band gap are lower than the band gap of bulk crystalline NiO, which is about 4.0 eV [42, 43]. This is due to the dopant Li ions, which act as scattering centers and affect on the carrier mobility [39].

*3. 4. Electrical properties*

The electrical resistivity values for NiO and NiO:Li films with 2% and 8% doping concentrations are displayed in Table 1. The resistivity decreases systematically with increasing lithium concentration. Undoped NiO is known as a p-type semiconductor, in which the p-type electrical conductivity is associated with the presence of interstitial oxygen atoms and nickel vacancies, which result in the formation of $Ni^{3+}$ ions [44]. For doped NiO, $Ni^{3+}$ can also be formed by the incorporation of lithium atoms in NiO, as shown in the following equation [44, 45]:

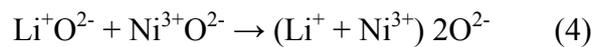

$$Li^+O^{2-} + Ni^{3+}O^{2-} \rightarrow (Li^+ + Ni^{3+})\, 2O^{2-} \qquad (4)$$

Lithium incorporation leads to an increase of the concentration of $Ni^{3+}$, which explains the decrease of the resistivity of the material after doping.

*3.5. Hydrogen sensing*

Gas sensitivity of NiO films is strongly related to adsorption/desorption processes occurring at the gas-film interface. The response $S$ of the NiO and NiO:Li samples to hydrogen is calculated according to Eq. (1). The electric resistance of NiO increases in the presence of hydrogen. This is because hydrogen reacts with oxygen, which is adsorbed on the surface of NiO, resulting in the release of $H_2O$ vapor and free electrons. The free electrons recombine with holes, which are the majority charge carriers in NiO. Therefore, the reduction in the number of majority charge carriers results in a resistance increase in the material [46]. This is typical for a p-type semiconductor, such as NiO, in the presence of a reducing gas, such as hydrogen. In this work, the effect of lithium doping on the hydrogen sensing properties of NiO (sensor response and operating temperature) was examined.

*3.5.1. Response to hydrogen for undoped NiO*

Undoped NiO was tested as a hydrogen sensor at an operating temperature of 180 °C for the gas concentration range of 15 000 – 1000 ppm. A typical response of this sample to hydrogen is shown in Fig. 5. The response of the films to hydrogen was rather slow, on the order of a few minutes, and slightly decreased with decreasing hydrogen concentrations. We notice, however, that the sensing response is not proportional to the gas concentration.

Rapid response and recovery times are requirements for most gas sensors. Figure 6 shows the response curves for the NiO sample under 10 000 ppm of hydrogen for different operating temperatures. In the presence of hydrogen, the response of the sample increases. From Fig. 6 we observe that the response is highest at 180 °C and the response time decreases with increasing operating temperature. The response time is defined as the time interval between 10% and 90% of the total signal change. The improvement in the sensor response time can be attributed to the high thermal energy of hydrogen molecules, which react with the adsorbed oxygen molecules.

Figure 7 shows the maximum NiO sensor response to hydrogen concentrations in air, ranging from 1000 to 15 000 ppm, at different operating temperatures. The response value increases with increasing concentration of the gas. The best response values are obtained at 180 °C for the undoped NiO sample. The response of the sensor increases with increasing operating temperature, because the adsorption–desorption kinetics, which affect the sensor performance, depend on the operating temperature [47, 48]. This sample did not respond at all to hydrogen concentrations lower than 10 000 ppm at 140°C.

*3.5.2. Response to hydrogen for Li-doped NiO*

Figures 8a, b, and c show the maximum response of NiO:Li sensors to different hydrogen concentrations in air for different operating temperatures. The response is clearly enhanced by the incorporation of lithium in the NiO thin films. At the same operating temperature, the response of NiO:Li sensors exhibits a clear increase in comparison to that of the undoped NiO sensor. The sensor response also increases with the lithium content of the NiO:Li films. The response difference between doped and undoped films increases for higher hydrogen concentrations. The increase in the response induced by lithium doping of NiO can be attributed to the change in the morphology of the doped films, in particular the change of the grain size and shape after adding lithium, which increases the active surface of the samples. The increase of the surface roughness leads to an increase of the active surface of the films, resulting in an increase of the available adsorption sites.

## 4. Conclusions

In this work, resistive NiO gas sensors were deposited on glass substrates by the sol-gel spin coating technique. Three different sensors were fabricated, using an undoped NiO film and two Li-doped NiO films with 2% and 8% lithium concentrations. According to X-ray diffraction measurements, the films are polycrystalline with a structure corresponding to the cubic structure of NiO, independent of the lithium dopant. The average crystallite size slightly decreases from 21 nm for undoped NiO to 18 nm by the incorporation of 8% lithium in the NiO film. On the other hand, the RMS roughness, estimated from AFM analysis, largely increases from 3 nm for the undoped NiO films to 13.5 nm for the doped NiO film with 8% lithium content. Lithium doping improves the sensor response to hydrogen. We attribute the increase in the sensor response to the increase of the surface roughness of the films with increasing lithium content.

## Acknowledgments

This work was supported by the Tunisian Ministry of Higher Education and Scientific Research. X-ray diffraction data in this work were acquired with an instrument supported by the unit x-ray diffraction, geology department, Faculty of sciences Tunis (FST). We thank Mr. K. Nasri and A. Ounis (FST) for the assistance in XRD data acquisition.

**Figure captions**

**Fig. 1:** X-ray diffraction patterns for NiO, NiO:Li 2%, and NiO:Li 8% samples.

**Fig. 2:** Three-dimensional (3D) AFM images of (a) NiO, (b) NiO:Li 2%, and (c) NiO:Li 8% samples.

**Fig. 3:** Grain size distribution of (a) NiO, (b) NiO:Li 2%, and (c) NiO:Li 8% samples, obtained from AFM measurements.

**Fig. 4:** Optical transmittance spectra and Tauc plot for NiO, NiO:Li 2%, and NiO:Li 8% samples.

**Fig. 5:** Response of the undoped NiO sensor to hydrogen at 180 °C operating temperature, for the hydrogen concentration range 15 000 – 1000 ppm.

**Fig. 6:** Response of the undoped NiO sensor to 10 000 ppm of hydrogen for different operating temperatures.

**Fig. 7:** Maximum response of the undoped NiO sensor to different hydrogen concentrations for different operating temperatures.

**Fig. 8:** Maximum response of NiO, NiO:Li 2%, and NiO:Li 8% sensors to different hydrogen concentrations for an operating temperature of (a) 140 °C, (b) 160 °C, and (c) 180 °C.

**Table caption**

**Table 1:** Structural, optical, and electrical properties of the NiO and NiO:Li films.

## Table 1

| Sample | Crystallite size D (nm) | Surface roughness (RMS) (nm) | Band gap energy $E_g$ (eV) | Resistivity $10^4$ ($\Omega$.cm) |
|---|---|---|---|---|
| **NiO** | 21 | 3 | 3.88 | 9.62 |
| **NiO:Li 2%** | 19 | 5 | 3.94 | 5.61 |
| **NiO:Li 8%** | 18 | 13.5 | 3.97 | 3.57 |

## Figure 1

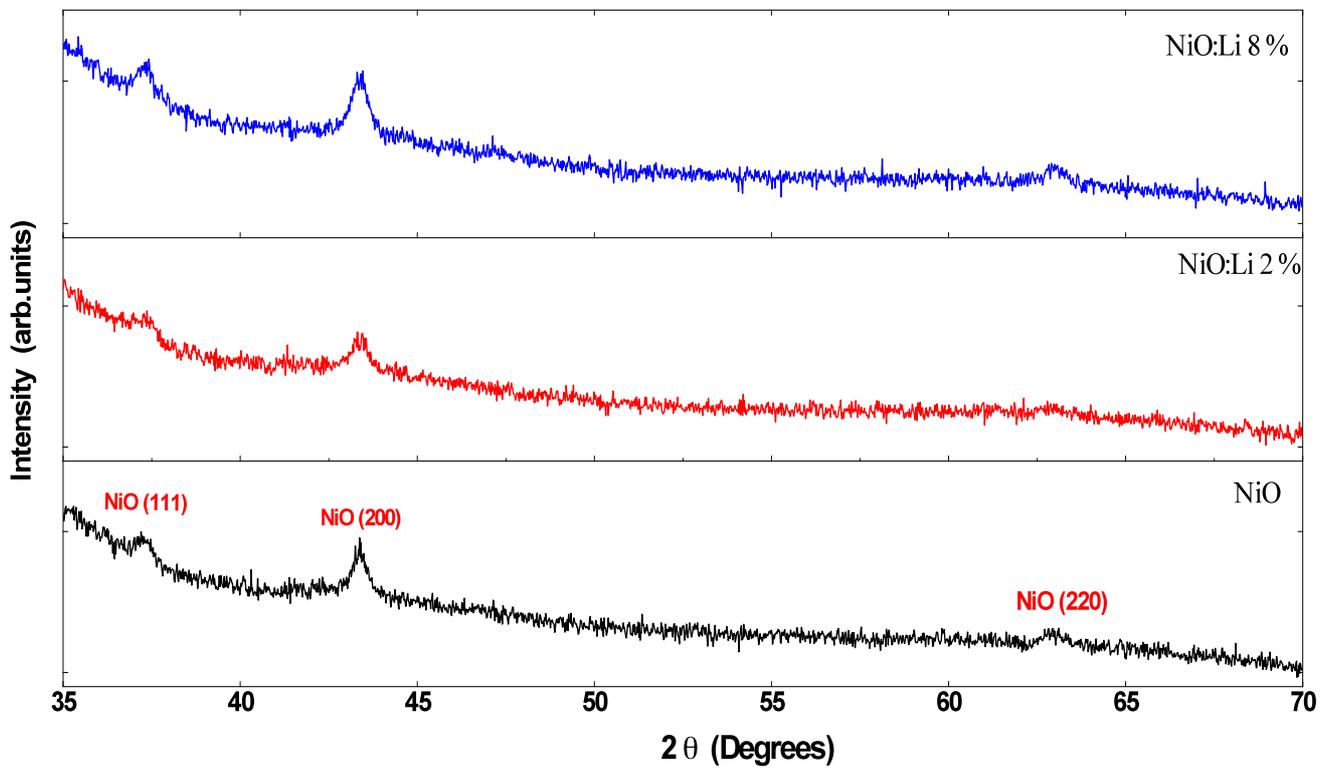

**Figure 2**

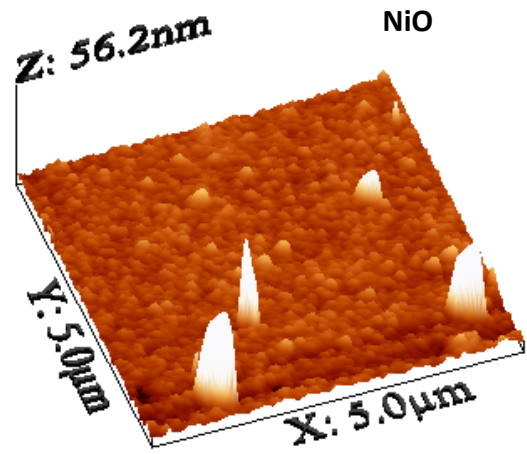

NiO

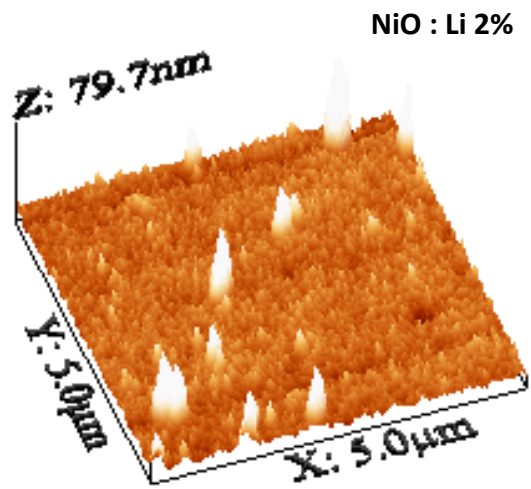

NiO : Li 2%

NiO : Li 8%

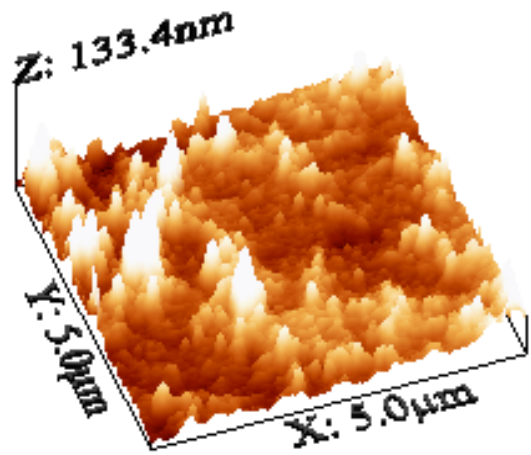

Figure 3

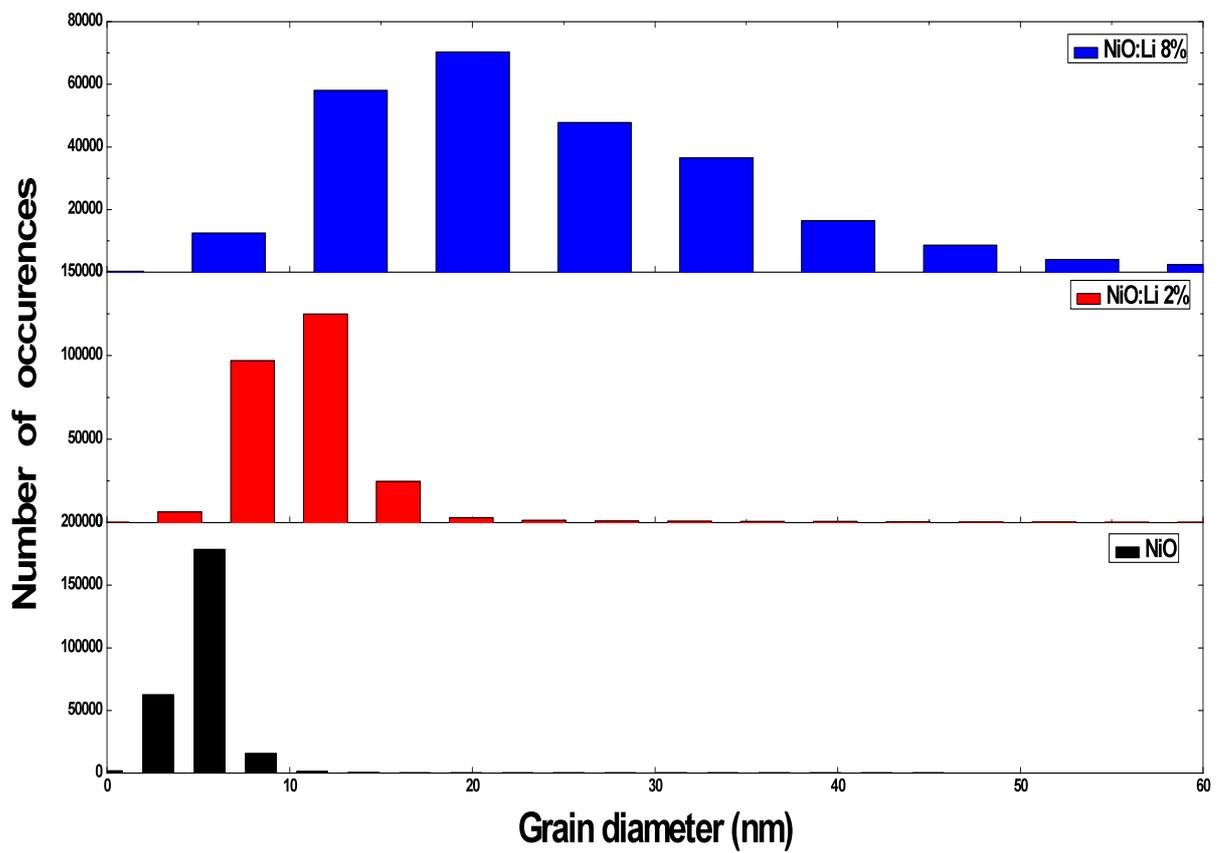

**Figure 4**

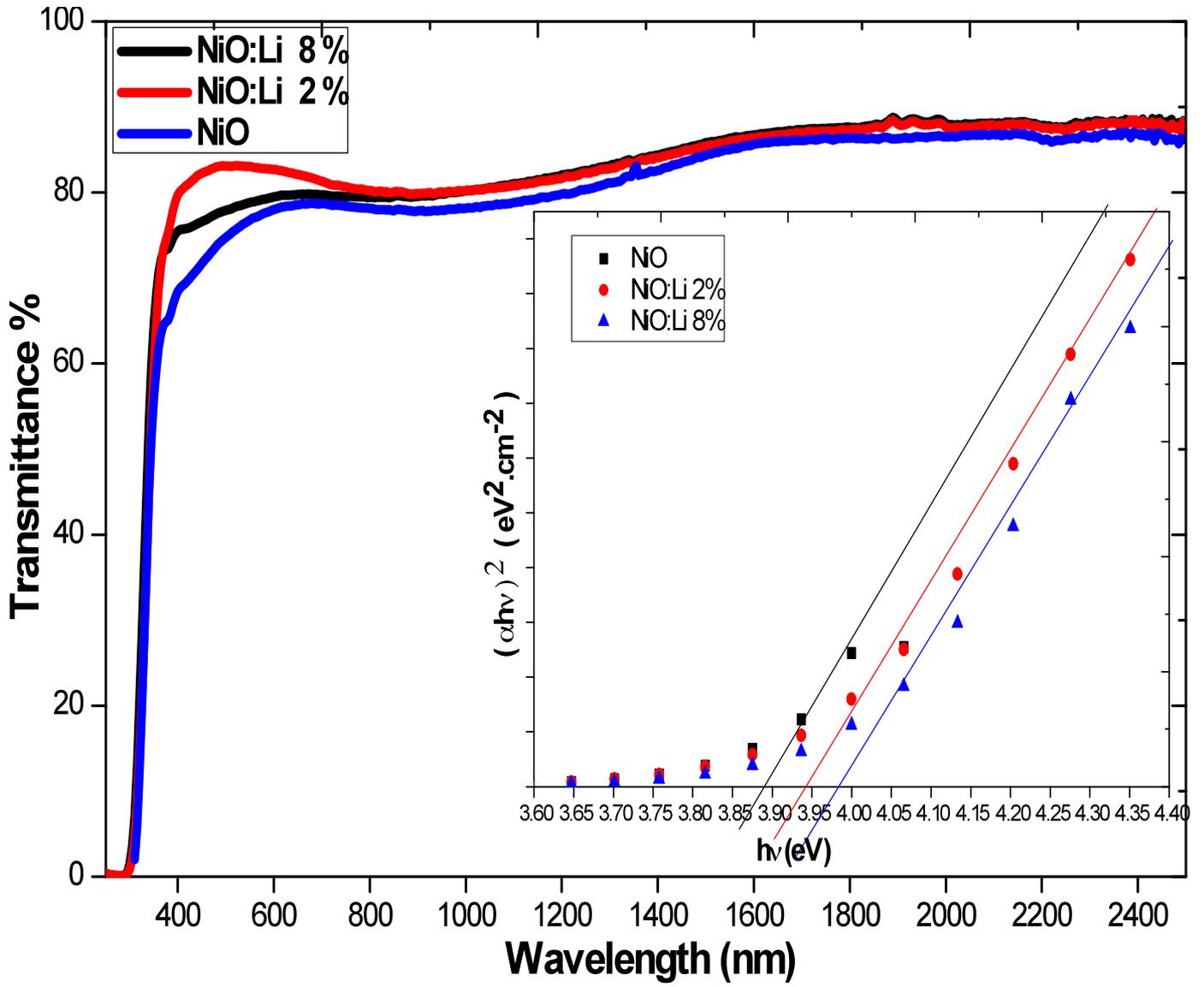

**Figure 5**

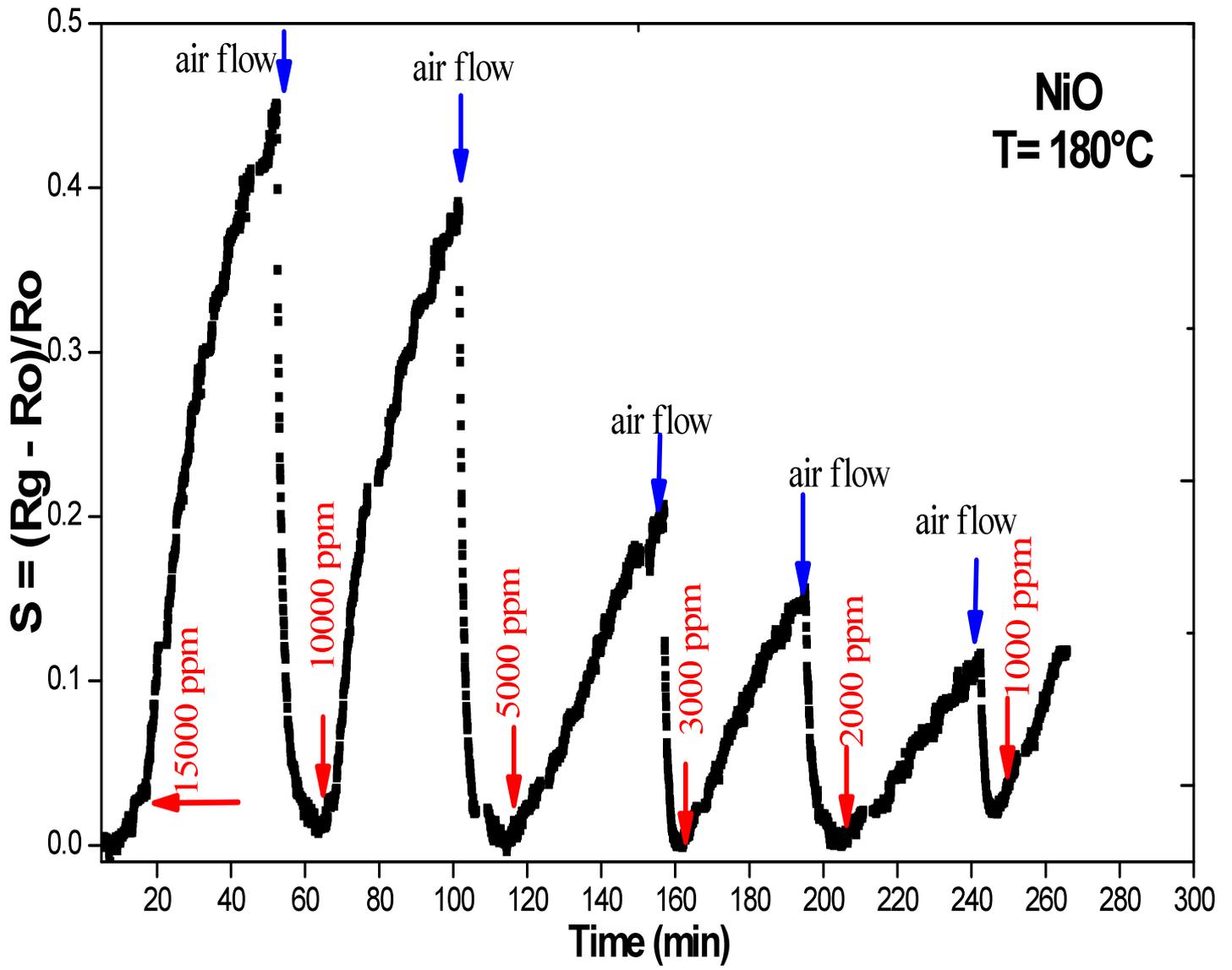

**Figure 6**

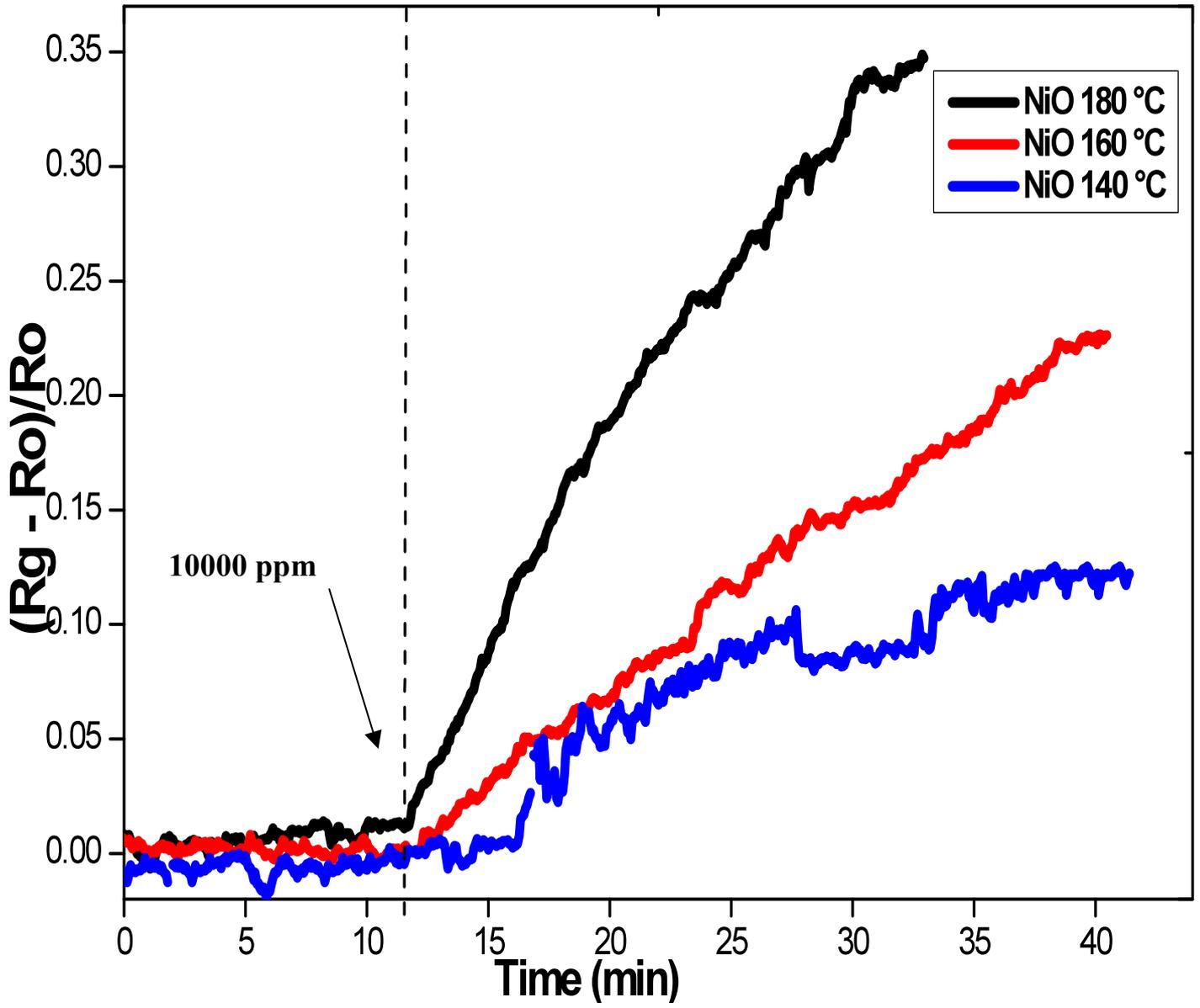

**Figure 7**

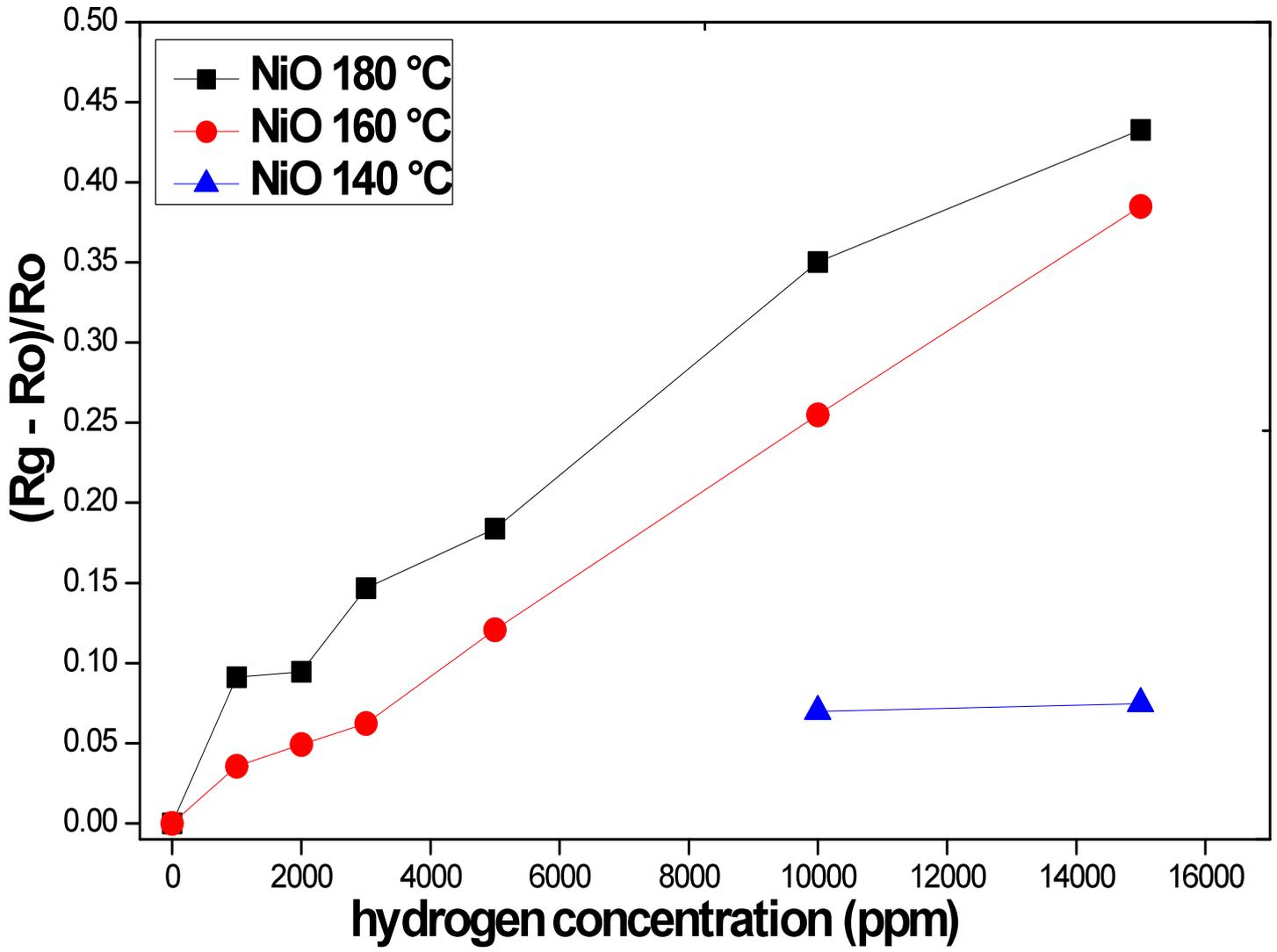

**Figure 8**

(a)

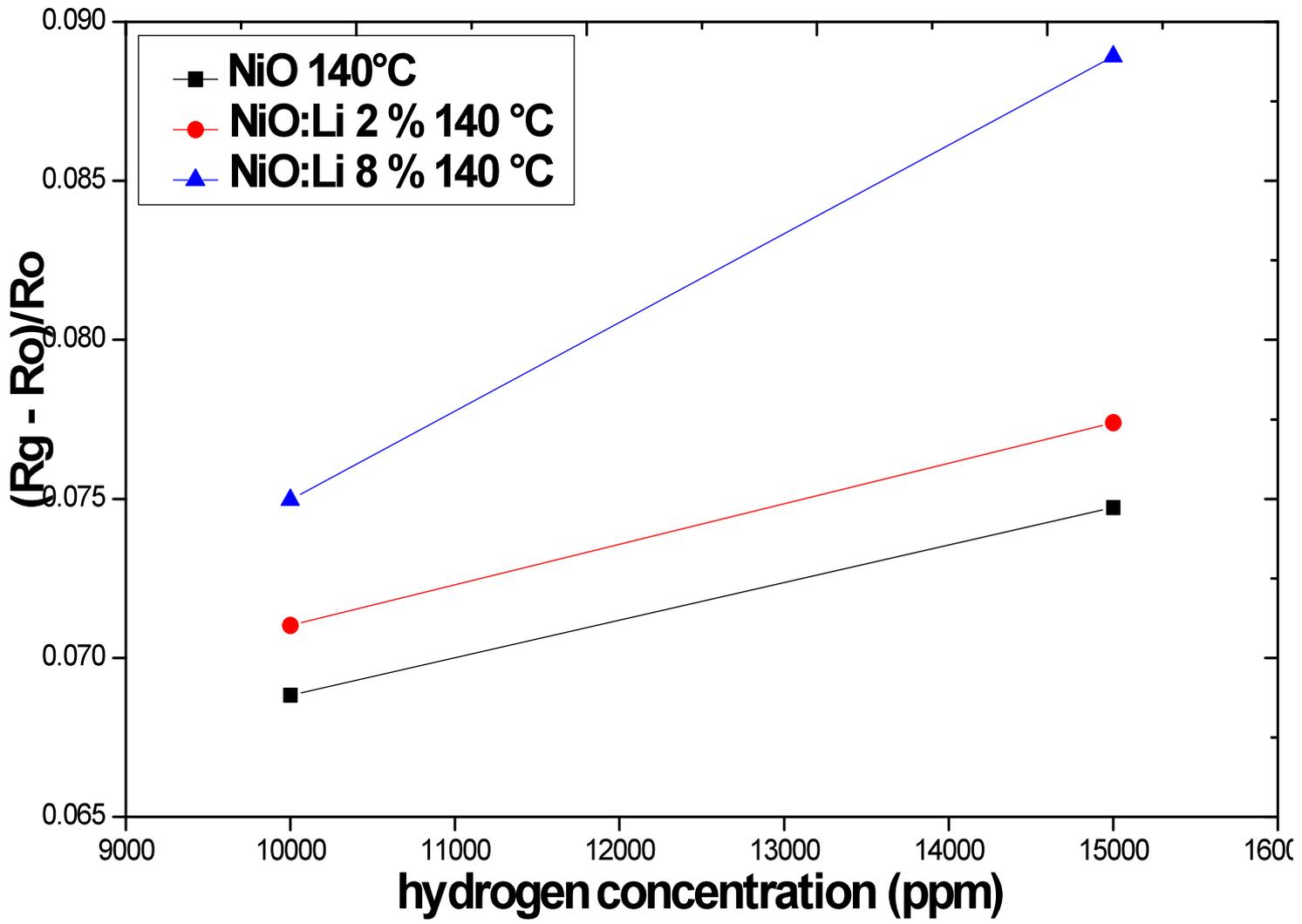

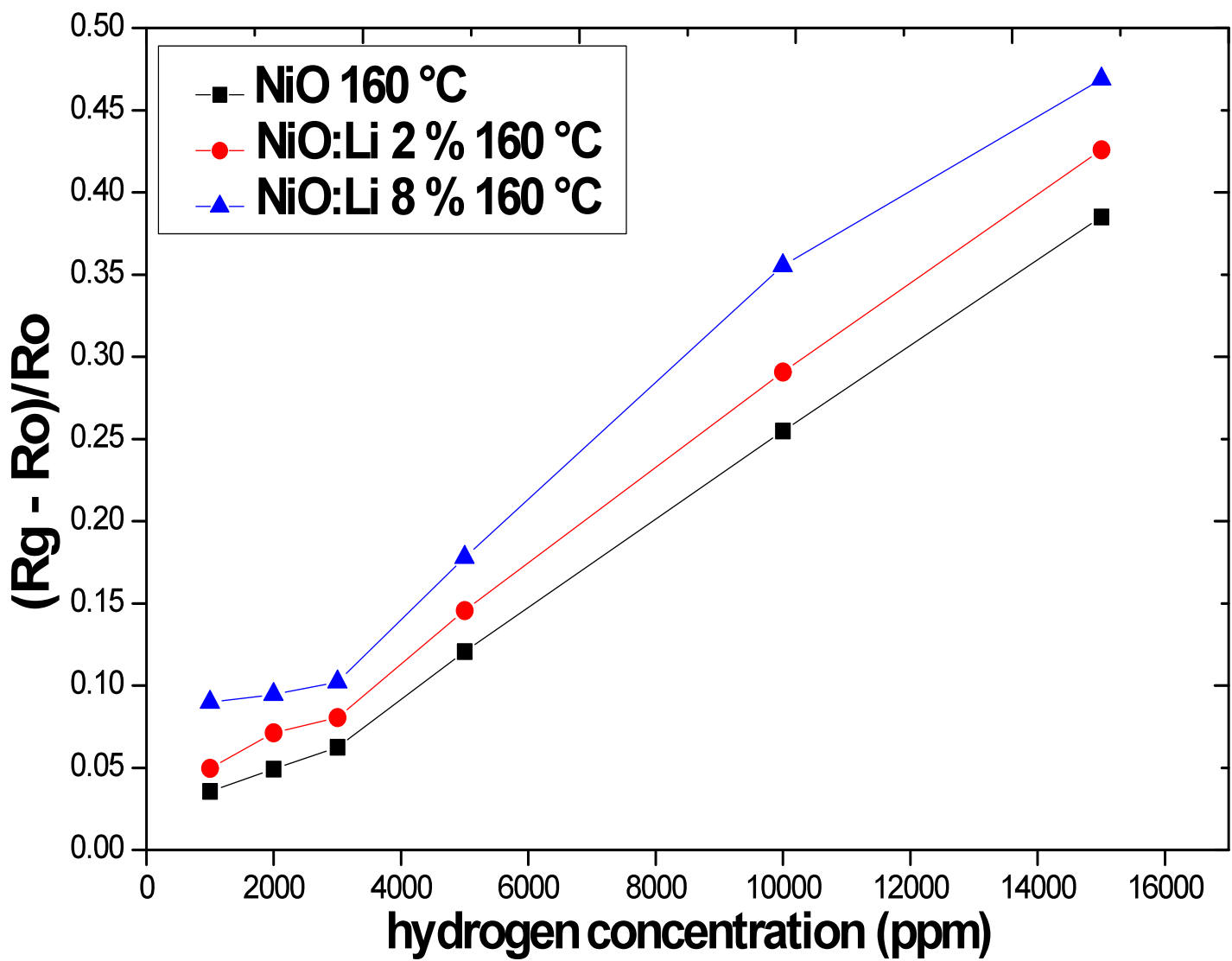

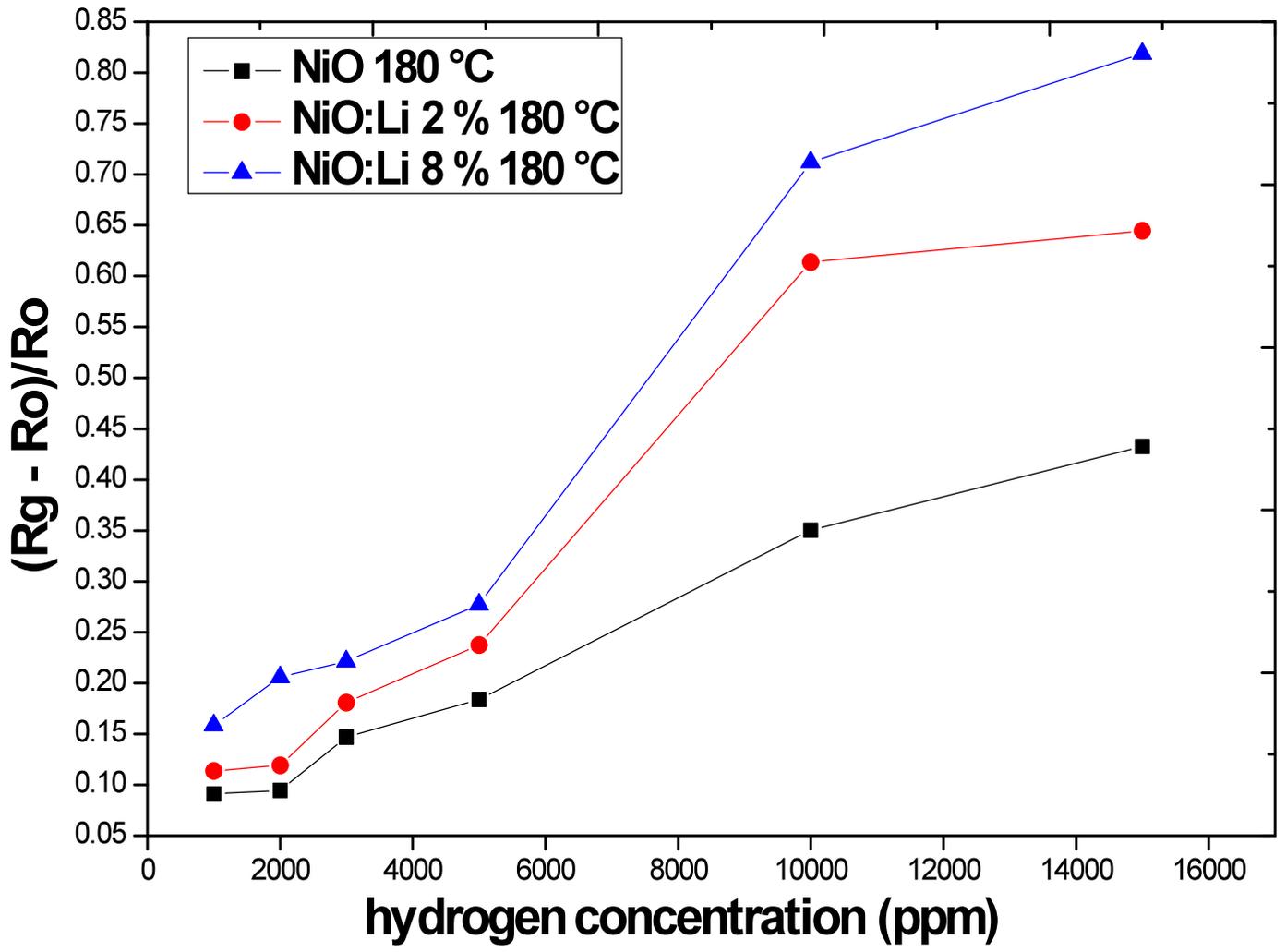